\documentclass[envcountsame,final]{llncs}
\usepackage[utf8]{inputenc}
\usepackage{misc0}

\usepackage{booktabs}
\usepackage{xspace}
\usepackage{amsmath}
\usepackage{amssymb}
\usepackage[obeyFinal]{todonotes}
\usepackage{cite}

\newcommand{\m}{\ensuremath{\mathsf{m}}\xspace}
\newcommand{\match}{\ensuremath{\mathsf{match}}\xspace}

\newcommand{\pair}[2]{\ensuremath{({#1},{#2})}}

\newcommand{\semiring}{\ensuremath{\mathbb{K}}\xspace}

\newcommand{\semiringVariables}{\ensuremath{\NV}\xspace}

\newcommand{\plus}{\ensuremath{+}\xspace}

\newcommand{\wt}{\ensuremath{\mathsf{wt}}\xspace}

\title{An Upper Bound for Provenance in \ELHr}
\author{Rafael Pe\~naloza}
\institute{University of Milano-Bicocca, Italy \\ \email{rafael.penaloza@unimib.it}}

\begin{document}
\maketitle

\begin{abstract}
We investigate the entailment problem in \ELHr
ontologies annotated with provenance information. 
In more detail, we show that subsumption entailment is in NP
if provenance is represented with polynomials from the Trio 
semiring and in PTime if the semiring is not commutative. 
The proof is based on the construction of a weighted tree automaton which recognises 
a language that matches with the corresponding  provenance polynomial.
\end{abstract}

\section{Introduction}

The study of provenance has recently gained interest in description logics as a manner to keep track of the sources that are
responsible for a consequence to follow from an ontology~\cite{bourgauxozaki,CalvaneseLantiOzakiPenalozaXiao19}. 
The basic idea behind provenance is to assign a unique label to each axiom in an ontology, and obtain a \emph{summary} of the
causes for deriving a consequence through two operators from a semiring: a \emph{product}, which combines together the
axioms used in one derivation, and a \emph{sum} which accumulates the products from the different possible derivations. These
two operations must satisfy some properties, forming a \emph{semiring}.

Although the motivation and the basic underlying structure is reminiscent
of axiom pinpointing \cite{BaPe10b,ScCo-03}, there are subtle but important differences which warrant further analysis.
A primary difference is that provenance does not require minimality of the information provided (as opposed to the notion of
\emph{justification}), but still requires a coherence between the provenance elements forming a so-called provenance \emph{monomial};
the \emph{product} of variables identifying the axioms needed to derive a desired consequence.
In addition, work in provenance is usually pursued in an abstract form, studying the properties based on a general semiring,
which can later be instantiated to specific algebraic structures depending on the application. Axiom pinpointing can indeed be obtained
by instantiating to a very specific semiring. 

Very recently, the problem of answering provenance queries in the description logic \ELHr was 
studied~\cite{DBLP:conf/ijcai/BourgauxOPP20}. That work focused on a semiring where the product operation is commutative
and idempotent, and expressed the provenance information through an expanded polynomial; that is, a sum of monomials. 
One of the main results was a consequence-based algorithm for the monomial of an entailment problem; that is, deciding whether
the provenance polynomial for a consequence contains a given monomial $m$. It was shown that this problem is in \textsc{PSpace},
but the best matching lower bound was the polynomial hardness for reasoning in \EL.

In this paper we improve that upper bound by showing that the monomial for a subsumption problem is in \textsc{NP}. To achieve
this goal, we view the completion algorithm from~\cite{DBLP:conf/ijcai/BourgauxOPP20} as a weighted tree automaton, which 
accepts all the completion-like proofs of a derivation. Through the behaviour of this automaton, the monomial problem is reduced 
to a membership problem in regular languages. To preserve polynomiality in the behaviour computation, we adapt the notion of
structure sharing to automata construction with the help of \emph{acyclic recursive automata} (also known as \emph{hierarchical
state machines} \cite{Yann-00}). These automata are exponentially more succinct than NFA, but not more expressive, and retain
most of the complexity properties of NFA.

\section{Preliminaries}

%\subsection{Semirings}
%\label{sec:sr}

A \emph{semiring} is an algebraic structure $\mathbb{S}=(S,\oplus,\otimes,\textbf{0},\textbf{1})$ where $\oplus$ and $\otimes$ are
associative binary operators over $S$ with neutral elements \textbf{0} and \textbf{1}, respectively, and such that $\oplus$ is
commutative, and $\otimes$ distributes over $\oplus$ \cite{Gola-92}. In the context of this paper, we consider two specific
well known semirings: the \emph{language} semiring, and the \emph{trio} semiring.

The \emph{language semiring} $\mathbb{L}=(\Lmc(\Sigma),\cup,\cdot,\emptyset,\{\varepsilon\})$ is the semiring of all languages
(that is, sets of finite words) over the alphabet $\Sigma$ with the usual concatenation of languages $\cdot$, and the union of sets
$\cup$. The empty word is denoted by $\varepsilon$. To reduce notation, we often represent singleton languages merely by the
word they contain, when it is clear from the context. 

The \emph{trio semiring} $\semiring=(  \mathbb{N}[\semiringVariables], \plus ,\times,0,1)$ 
is the semiring of polynomials with
coefficients in $\mathbb{N}$ and variables in a countably infinite set $\semiringVariables$, with 
the operation $\plus$ defined as usual and $\times$ idempotent and commutative~\cite{Green07-provenance-seminal,GreenT17,DBLP:journals/ftdb/CheneyCT09}.
We also consider 
in Section~\ref{sec:non-commutative} the case in which $\times$ is non-commutative.
Polynomials in the trio semiring  in this work are in \emph{expanded form}, meaning that  they 
are sums of monomials. %By distributivity, 
Every polynomial can be represented in this form.

Objects of the language and the trio semirings are similar, but have subtle differences: every language \Lmc
can be seen as a (potentially infinite) polynomial, where each monomial is a word in \Lmc. Conversely, the class of all monomials
in a polynomial $P$ in expanded form can be seen as a language modulo the commutativity of the product $\times$.

We consider a syntactic restriction of the ontology language \ELHr~\cite{dlhandbook}. 
Concept and role names are taken from the disjoint countable sets \NC and \NR, respectively, also
disjoint from \NV. 
\ELHr \emph{general concept inclusions} (GCIs)  $C\sqsubseteq D$ are   
built through
the grammar rules
$C::= A\mid \exists R.C \mid C\sqcap C\mid \top$,
$D::= A\mid\exists R$,
where $R\in \NR$,~$A\in \NC$.
\emph{Role inclusions} (RIs) and 
\emph{range restrictions} (RRs) are of the form 
$R\sqsubseteq S$ and ${\sf ran}(R)\sqsubseteq A$, respectively,
 with $R,S\in\NR$ and $A\in\NC$.
 An \ELHr \emph{axiom} is a GCI, RI, or RR. %, or assertion. 
An \ELHr TBox is a finite set of \ELHr axioms. 
The reason for syntactically restricting \ELHr 
is that conjunctions or qualified restrictions of a role on the right-hand side 
of GCIs lead to counter-intuitive behavior when adding provenance annotations; see~\cite{DBLP:conf/ijcai/BourgauxOPP20} for
a detailed discussion on this issue. 

An \emph{annotated \ELHr TBox} \Tmc is a set of 
\ELHr axioms, each annotated with an element from $\variable\in\NV\cup\{1\}$ representing 
provenance information.
Axioms annotated with provenance information can be derived from 
an annotated ontology. They are annotated with monomials (potentially with more than one variable) representing 
the derivation of the axiom w.r.t. \Omc. From now on, \NM represents the set of all monomials.

An \emph{annotated interpretation} is a triple
$\Imc=(\Delta^\Imc,\Delta^\Imc_{\sf m},\cdot^\Imc)$
where $\Delta^\Imc,\Delta^\Imc_{\sf m}$ are non-empty disjoint sets (the
\emph{domain} 
and 
\emph{domain of monomials} of \Imc, respectively),
and $\cdot^\Imc$ maps
\begin{itemize}
%\item every $a\in\NI$ to $a^\Imc\in\Delta^\Imc$;
\item   every $A\in\NC$ to $A^\Imc\subseteq \Delta^\Imc\times \Delta^\Imc_{\sf m}$;
\item every   $R\in\NR$ to
$R^\Imc\subseteq \Delta^\Imc\times\Delta^\Imc\times \Delta^\Imc_{\sf m}$; and
\item  every $\monomial,\nonomial\in \NM$
to $\monomial^\Imc,\nonomial^\Imc\in\Delta^\Imc_{\sf m}$
s.t.\ $\monomial^\Imc=\nonomial^\Imc$ iff $\monomial$ and $\nonomial$ are  equal
modulo associativity, commutativity and $\times$-idempotency 
(e.g., $(\nonomial\times \monomial)^\Imc = (\monomial\times \nonomial)^\Imc$).
\end{itemize}
As mentioned, we consider in Section~\ref{sec:non-commutative} the case in which $\times$ is non-commutative.
We extend
$\cdot^\Imc$ to complex \ELHr expressions as usual: 
  \begin{align*} 
	(\top)^\Imc = {} & \Delta^\Imc\times \{1^\Imc\}; \\ 
	(\exists R)^\Imc = {} & \{(d,\monomial^\Imc)\mid \exists e\in\Delta^\Imc 
	 \text{ s.t. }(d,e,\monomial^\Imc)\in R^\Imc\};\\
    (C\sqcap D)^\Imc = {} & \{(d,(\monomial\times \nonomial)^\Imc)\mid (d,\monomial^\Imc)\in C^\Imc, (d,\nonomial^\Imc)\in D^\Imc\}; \\
    ({\sf ran} (R))^\Imc = {} & \{(e,\monomial^\Imc)\mid \exists d\in\Delta^\Imc
    \text{ s.t. }(d,e,\monomial^\Imc)\in R^\Imc\};   \\ 
    (\exists R.C)^\Imc = {} & \{(d,(\monomial\times \nonomial)^\Imc)\mid \exists e\in\Delta^\Imc
    \text{ s.t. }\\
	    & \ \ (d,e,\monomial^\Imc)\in R^\Imc, (e, \nonomial^\Imc)\in C^\Imc\}. 
  \end{align*} 
The annotated interpretation \Imc \emph{satisfies}:
%\centerline{$
%  \begin{array}{ll}
    $\pair{R\sqsubseteq S}{\monomial}$ if for all 
    		$\nonomial\in\NM, (d,e,\nonomial^\Imc)\in R^\Imc$ implies $(d,e,(\monomial\times \nonomial)^\Imc)\in S^\Imc\!$;  and
    $\pair{C\sqsubseteq D}{\monomial}$ if for all 
    		$\nonomial\in\NM, (d,\nonomial^\Imc)\in C^\Imc$ implies $(d,(\monomial\times \nonomial)^\Imc)\in D^\Imc\!$.
%   	\\
%\pair{A(a)}{\monomial}, & \text{if } (a^\Imc,\monomial^\Imc)\in A^\Imc\!; \quad \text{ and } \\ 
%   \pair{R(a,b)}{\monomial}, & \text{if } (a^\Imc,b^\Imc,\monomial^\Imc)\in R^\Imc\!.  	
%  \end{array}
%$}
\Imc is a \emph{model} of \Tmc, denoted $\Imc\models\Tmc$, if
it satisfies all annotated axioms in \Tmc.   
$\Tmc$ \emph{entails} $(\alpha,\monomial)$, denoted $\Omc\models (\alpha,\monomial)$,
if 
$\Imc\models (\alpha,\monomial)$ for every model $\Imc$ of $\Omc$.  

We are interested in the provenance for a subsumption problem: given a TBox \Tmc and two concept names $A,B$, find
all monomials $m$ such that $\Tmc\models (A\sqsubseteq B, m)$. We solve it by constructing an ARA that
accepts representatives of all these monomials. We use this construction to answer, given a monomial $m$,
whether $\Tmc\models (A\sqsubseteq B,m)$ holds, and show that this problem is in \textsc{NP}.

\section{Automata}

We consider two generalisations of non-deterministic finite automata (NFA) \cite{HoUl-79automata}; 
namely, weighted tree automata and acyclic recursive automata.

\subsection{Weighted Tree Automata}

Tree automata \cite{tata2007} generalise NFAs by accepting trees rather than words; the branching of the trees is identified by 
the \emph{arity}
of the automaton. Weighted tree automata further generalise this notion by not only accepting or rejecting an input tree, but assigning
a value from a given semiring (its \emph{weight}) \cite{HWA09}. For the scope of this paper, we consider only \emph{unlabelled} trees.

Let $k\ge 1$. A \emph{weighted tree automaton} over $\mathbb{S}$ of arity $k$ is a tuple of the form 
$\Amc=(Q,\mathbb{S},\wt,I,f)$ where $Q$ is a finite set of \emph{states}, $\mathbb{S}=(S,\oplus,\otimes,\textbf{0},\textbf{1})$ is 
a semiring,
$\wt:Q^{k+1}\to S$ is the \emph{transition weight function}, $I\subseteq Q$ is the set of \emph{initial states}, and
$f:Q\to S$ is the \emph{exit weight function}. 

As usual, we represent trees of arity $k$ as finite non-empty sets $T\subseteq\{1,\ldots,k\}^*$ such that if $wi\in T$, then $w,wj\in T$
for each $w\in\{1,\ldots,k\}^*, 1\le i,j\le k$.
Given a tree $T$ of arity $k$, a \emph{run} $\rho:T\to Q$ of \Amc over $T$ assigns a state to each node in $T$. The \emph{weight} 
of this run is $\wt(\rho)=\bigotimes_{w1\in T}\wt(\rho(w),\rho(w1),\ldots,\rho(wk))\otimes\bigotimes_{w1\notin T}f(\rho(w))$; that is, the 
product of all the transition and exit weights given the states assigned by $\rho$. For non-commutative semirings, this product is 
made from the root to the leafs, and in the order of the children (i.e., top-down, left-to-right).
Given a state $q\in Q$, we define $\wt(q)=\bigoplus_{\rho(\varepsilon)=q}\wt(\rho)$; that is, the sum of the weights of all runs
that label the root of a tree with $q$. The \emph{behaviour} of the automaton \Amc is the sum of the weights of all its initial states
$\|\Amc\|:=\bigoplus_{q\in I}\wt(q)$.

\subsection{Acyclic Recursive Automata}

Acyclic recursive automata generalise NFA by allowing an automaton to call another one, but the calls between automata must 
respect a hierarchical ordering. They were originally introduced as \emph{hierarchical state machines} \cite{Yann-00} with a 
slightly different structure. 

\begin{definition}[ARA]
An \emph{acyclic recursive automaton} over the alphabet $\Sigma$ is a finite set $\Amf=\{\Amc_i\mid i\in I\}$ of NFAs
$\Amc_i=(Q_i,\Sigma_i,\Delta_i,I_i,F_i)$, where $(I,\le)$ is a partially ordered set of indices, such that:
%\begin{itemize}
(i) for all $i\not=j\in I$, $Q_i\cap Q_j=\emptyset$;
(ii) $\Sigma_i=\Sigma\cup \{\m_j\mid j<i\}$; and
(iii) $\{\m_i\mid i\in I\}\cap \Sigma=\emptyset$.
%\end{itemize}
\end{definition}
We call the symbols $\m_i$, which are added to the alphabets of the different automata in \Amf, \emph{call triggers}.
When an automaton $\Amc_i$ reads the symbol $\m_j$, it ``calls'' the automaton $\Amc_j$, which continues reading
the word until it chooses to return the control to the ``calling'' automaton $\Amc_i$ (signalled by the symbol $\overline\m_j$). 
This return is only possible if $A_j$ is in one
of its accepting states. In practice, the automaton $\Amc_j$ is in charge of accepting a portion of the input word. 

Each automaton $\Amc_i$ may call any other automaton $\Amc_j$ where $j<i$. Hence, there
may be a sequence of nested calls, but the depth of this nesting is always bounded by the number $n$ of automata in \Amf.
Moreover, the automaton $\Amc_i$ can never call itself either directly or indirectly.
To define the language accepted by the ARA \Amf, we adapt the notion of a run to take into account also
the nested calls between the automata.

\begin{definition}[valid run]
A \emph{run} of the ARA $\Amf=\{\Amc_j\mid j\in I\}$ is a finite sequence $\rho=q_0,s_1,q_1,\ldots,s_k,q_k$
such that $q_i\in\bigcup_{j\in I} Q_j$ for all $0\le i\le k$ and $s_i\in\Sigma\cup\{\m_j,\overline \m_j\mid j<i\}$ for all
$1\le i\le k$.
The notion of a valid run on an automaton $\Amc_i$ is inductively defined as follows. 
The run $\rho=q_0,s_1,q_1,\ldots,s_k,q_k$ is \emph{valid} on $\Amc_i$ iff
\begin{itemize}
\item $\{s_1,\ldots,s_k\}\subseteq \Sigma$ and $(q_j,s_{j+1},q_{j+i})\in\Delta_i$ for all $0\le j<k$ or
\item $j$ is the smallest index such that $s_j\notin\Sigma$, $s_j$ is of the form $\m_\ell$, there exists $j'>j$ with 
	$s_{j'}=\overline \m_\ell$, and 
	\begin{itemize}
		\item $q_0,s_1,q_2,\ldots,s_{j-1},q_{j-1}$ and $q_{j'},s_{j'+1},\ldots,q_k$ are valid in $\Amc_i$
		\item $q_j,s_{j+1},q_{j+1},\ldots,s_{j'-1},q_{j'-1}$ is valid in $\Amc_\ell$ and
		\item $(q_{j-1},\m_\ell,q_{j'})\in \Amc_i$
	\end{itemize}
\end{itemize}
\end{definition} 
So far, we have not yet expressed the use of initial and final states in accepting a word. In a nutshell, whenever we call an automaton, 
its execution should accept a segment of the input word, by traversing from an initial to a final state.

\begin{definition}[successful run]
Given a valid run $\rho=q_0,s_1,q_1,\ldots,s_k,q_k$, the index $i, 1\le i\le k$ is called a \emph{top-level call index} iff
$s_i=\m_j$ for some $j$, and for every $\ell<i$ such that $s_\ell=\m_{j'}$ there is an $\ell', \ell<\ell'<i$ such that 
$s_{\ell'}=\overline \m_{j'}$. If $i$ is a top-level call index with $s_i=\m_j$, then the smallest index $\ell,i<\ell\le k$ such that
$s_\ell=\overline \m_j$ is its \emph{match}. This is denoted as $\ell=\match(i)$.

A valid run $\rho=q_0,s_1,q_1,\ldots,s_k,q_k$ is \emph{successful} in $\Amc_i$ iff $q_0\in I_i$, $q_k\in F_i$ and for every
top-level call index $i$ with $s_i=\m_j$ and $\ell=\match(i)$, the sequence $q_{i},s_{i+1},\ldots,q_{\ell-1}$ is a successful
run in $\Amc_j$.

The run $\rho$ is successful in the ARA $\Amf=\{\Amc_i\mid i\in I\}$ iff it is successful in $\Amc_j$, for some maximal element
$j$ of $I$. The word 
\emph{accepted} by this run is the concatenation of all symbols of $\Sigma$ appearing in $\rho$. The \emph{language} of
\Amf is the set $\Lmc(\Amf)$ of all words accepted by a successful run in \Amf. By extension, the language accepted by $\Amc_i$
is the set $\Lmc(\Amc_i)$ of all words accepted by a successful run in $\Amc_i$, for each $i\in I$.
\end{definition}
ARAs are not more expressive than NFAs; they also accept regular languages. The main difference is that an ARA can
be exponentially more succinct than an NFA for representing a given language. For example, the language $\{a^{2^n}\}$ that
contains only one word with $2^n$ symbols $a$ can only be recognised by NFAs with at least $2^n$ states, but is 
accepted by an ARA having $n$ automata with 3 states each (hence $3n$ states in total); see Appendix \ref{app:succ}. 
The \emph{size} of the ARA \Amf is the total number of states in the NFAs in \Amf.

The relevant properties of ARAs for this paper are the following. Deciding whether the ARA \Amf accepts a word $w$ requires
only polynomial time. The concatenation of $n$ ARAs is obtained by adding a new NFA with $n+1$ states that calls each ARA
once. The union of $n$ ARAs is obtained by adding a new NFA with 2 states, which non-deterministically calls one of the ARAs.
Abusing the notation, given two ARAs \Amf, \Bmf we denote as $\Amf\cdot\Bmf$ and $\Amf\cup\Bmf$ the ARAs obtained 
through these constructions, respectively.

\section{The Weighted Automaton}

Our goal is to build an ARA which accepts representatives for all the monomials in the provenance of a subsumption relation. To do
so, we first present a weighted tree automaton whose behaviour (which is a language) can be seen as a polynomial (in expanded 
form) constructed by the provenance monomials for the desired consequence; modulo commutativity. 
The method for \emph{computing} this behaviour
will give rise to the ARA. 
%In the following, we will often equate words and languages with monomials and polynomials, respectively.
%The relationship between the terms should be obvious.\todo{I am not sure if we want to do this still. I leave the sentence here for
%now}

The construction of the automaton is based on considering the ``proofs'' of a derivation based on the consequence-based algorithm,
built in a top-down manner (from the desired consequence, deconstructed back to the axioms used). 
Formally, we have a different automaton for each consequence that we might want to verify. However, all the automata are 
equivalent, except for the initial state; which refers to the desired consequence. The automaton, which reads trees of 
arity 5, is also very simple because all transitions that refer to a consequence step have weight $\{\varepsilon\}$ (the neutral of
the language semiring product), and the only ``real''
weight is found at the final states (the exit weight) which is given by the provenance label of the axiom in the TBox.

Let $A_0, B_0$ be two distinguished concept names appearing in an annotated TBox \Tmc; the weighted automaton 
$\Amc_{A_0\sqsubseteq B_0}=(Q\cup\{\Box\}, \mathbb{L}, \wt, I, f)$ is given 
by
\begin{itemize}
\item $Q$ is the set of all axioms in restricted normal form on the alphabet of \Tmc;
\item $\wt(\delta)=\{\varepsilon\}$ if $\delta\in T$ (see Table \ref{tab:transA}) and $\wt(\delta)=\emptyset$ otherwise;
\item $I=\{A_0\sqsubseteq B_0\}$;
\item $f(q)=\{v\}$ if $(q,v)\in\Tmc$; $f(q)=\{\varepsilon\}$ if $q\in\{X\sqsubseteq X,X\sqsubseteq\top,\Box\}$; 
	and $f(q)=\emptyset$ otherwise.
\end{itemize}
\begin{table}[tb]
\caption{Transitions of $\Amc_{A_0\sqsubseteq B_0}$ with weight $\{\varepsilon\}$.}
\label{tab:transA}
\vspace*{-6mm}
\begin{align*}
T =  \{ & 
		(R_1 \sqsubseteq R_3, R_1 \sqsubseteq R_2, R_2 \sqsubseteq R_3, \Box, \Box, \Box), \\ &
		(\ran(R)\sqsubseteq A, R\sqsubseteq S, \ran(S)\sqsubseteq A, \Box, \Box, \Box), \\ &
		(A \sqsubseteq \exists S, A \sqsubseteq \exists R, R\sqsubseteq S, \Box, \Box, \Box),
		(A \sqsubseteq C, A \sqsubseteq B, B \sqsubseteq C, \Box, \Box, \Box), \\ &
		(A \sqsubseteq \exists R, A \sqsubseteq B, B \sqsubseteq \exists R, \Box, \Box, \Box), \\ &
		(A \sqsubseteq C, A \sqsubseteq B_1, A \sqsubseteq B_2, B_1\sqcap B_2\sqsubseteq C, \Box, \Box), \\ &
		(\ran(R) \sqsubseteq C, \ran(R)\sqsubseteq B_1, \ran(R)\sqsubseteq B_2, B_1\sqsubseteq C_1, B_2\sqsubseteq C_2,
			C_1\sqcap C_2\sqsubseteq C), \\ &
		(A \sqsubseteq C, A\sqcap B\sqsubseteq C, \top\sqsubseteq B, \Box, \Box, \Box), \\ &
		(A \sqsubseteq D, A \sqsubseteq \exists S, \ran(S)\sqsubseteq B, B\sqsubseteq C, S\sqsubseteq R, 
			\exists R.C\sqsubseteq D), \\ &
		(A \sqsubseteq C, A \sqsubseteq \exists R, \top \sqsubseteq B, \exists R.B\sqsubseteq C,\Box, \Box) \\ & \qquad
		\mid A,B,C,D\in N_C(\Tmc)\cup\{\top\}, R,S\in N_R(\Tmc) \}
\end{align*}
\end{table}
The special symbol $\Box$ is used to keep the arity of the automaton to $5$. In a nutshell, the transitions of this automaton
can be seen as the completion rules from~\cite{DBLP:conf/ijcai/BourgauxOPP20}, but applied \emph{backwards}, from the
consequence to the premises that generate it. The weight of any run which labels the root with $A_0\sqsubseteq B_0$ is either
$\emptyset$ if the labelled tree does not represent a derivation of the consequence, or a single word concatenating the annotations
of the axioms from \Tmc used in the derivation. Modulo idempotency, this word represents a provenance monomial for 
$A_0\sqsubseteq B_0$. The behaviour of the automaton is then the language containing a representation of all such provenance
monomials.
\begin{example}
\label{exa:treerun}
Consider the annotated TBox \Tmc containing the following five axioms
$\Tmc:=\{(B\sqcap C\sqsubseteq D,u), (\top \sqsubseteq B,v), (A \sqsubseteq C, w), (A \sqsubseteq \exists R, x),
(\exists R.B\sqsubseteq B,y)
\}$.
One possible run of the automaton $\Amc_{A\sqsubseteq D}$ is depicted in Figure~\ref{fig:treerun}.
\begin{figure}[tb]
\centering
\includegraphics{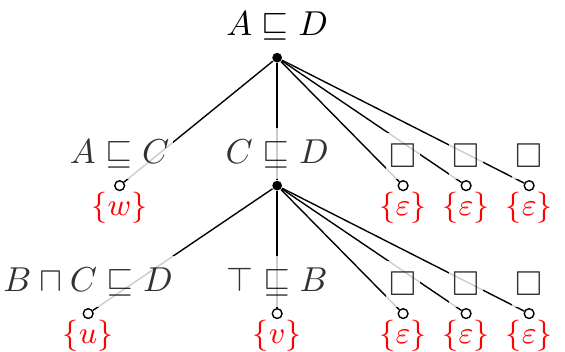}
\caption{One run of the tree automaton $\Amc_{A\sqsubseteq D}$ from Example \ref{exa:treerun}. Its weight is $\{wuv\}$.}
\label{fig:treerun}
\end{figure}
The two internal nodes (marked with \tikz \node[circle,fill,inner sep=1pt] {};) have a transition weight of $\{\varepsilon\}$.
The weight of this run is $\{wuv\}$. It can be seen that every non-leaf node is the consequence obtained from its successors
(ignoring the dummy nodes $\Box$). There are at least two other runs with weight different from $\emptyset$; one has weight
$\{vwu\}$ and the other $\{xvywu\}$. The behaviour $\|\Amc_{A\sqsubseteq D}\|$ contains $\{wuv,vwu,xvywu\}$. Note that the
first two words correspond to the same provenance monomial in \semiring, due to commutativity. In $\mathbb{L}$, they are two
different objects.
\end{example}
As it can be seen from the example, the behaviour of $\Amc_{A_0\sqsubseteq B_0}$ does not directly yield the set of all
provenance monomials for $A_0\sqsubseteq B_0$. However, these monomials can be extracted from $\|\Amc_{A_0\sqsubseteq B_0}\|$
by taking into account the commutativity and idempotency of \semiring. Abusing the notation, given a word $\omega$, we will
denote as $[\omega]$ a representative monomial $\omega$ w.r.t.\ commutativity and idempotency. Hence for instance
$[xuxv]=uvx$.
The next theorem is a direct consequence of the correctness of the completion algorithm \cite{DBLP:conf/ijcai/BourgauxOPP20}.
\begin{theorem}
\label{thm:aut:prov}
There is a run $\rho$ of $\Amc_{A_0\sqsubseteq B_0}$ with weight $\wt(\rho)=\{m\}\not=\{\varepsilon\}$ 
iff $\Tmc\models (A_0\sqsubseteq B_0, [m])$.
\end{theorem}
Thus, the behaviour of this automaton, which accumulates the weights of all possible runs, represents all provenance monomials for the
consequence $A_0\sqsubseteq B_0$. The question is: how to find this behaviour? Answering this question is the scope of the
following section; but before that, we emphasise that the automata for different consequences are all identical except for the initial
state, which is used to label the root node (that is, the goal that we aim to reach through a proof). Hence, for the TBox in Example
\ref{exa:treerun}, we get $\{v,xvy\}\subseteq\|\Amc_{A\sqsubseteq B}\|$.

\section{The Behaviour}
\label{sec:behaviour}

Following the general idea from \cite{BaPe10,DrKR-FI08}, we compute the behaviour of the automaton via a bottom-up approach, 
by iteratively accumulating the provenance of intermediate
consequences used in the derivation of $A_0\sqsubseteq B_0$. However, the technique must be adapted to handle the semiring
$\mathbb{L}$, which is not a lattice.

Specifically, we build the functions $\wt_i: Q\to \Lmc(N_V)$, $i\in\mathbb{N}$ as follows:
\begin{itemize}
\item $\wt_0 = f$;%
\footnote{Recall that $f$ is the exit weight function of the automaton.}
\item for $i\ge 0$, $\wt_{i+1}(q)=\wt_i(q)\cup \bigcup_{(q,q_1,\ldots,q_5)\in T}\wt_i(q_1)\cdot \cdots \cdot \wt_i(q_5)$
\end{itemize}
It can be shown by induction on $i$ that $\wt_i(q)$ has a representative for all the monomials arising from trees with root labelled
with $q$ and depth at most $i$. In particular for $i=0$, $\wt_0(q)$ is the label of the axiom $q$ if it appears in \Tmc, $\{\varepsilon\}$
if $q$ is a tautology or $\Box$, and $\emptyset$ otherwise. Importantly,
$\wt_i(q)\subseteq \wt_{i+1}(q)$ for all $q\in Q$ and all $i\in\mathbb{N}$. We can thus see the construction of $\wt_i$ as a 
monotone operator which, in particular, has a smallest fixpoint: the limit of the functions $\wt_i$. This fixpoint is, in fact, the behaviour
of $\Amc_{A_0\sqsubseteq B_0}$.

\begin{theorem}
\label{thm:behaviour}
The behaviour of $\Amc_{A_0\sqsubseteq B_0}$ is $\lim_{n\to\infty}\wt_n(A_0\sqsubseteq B_0)$.
\end{theorem}
Importantly, the functions $\wt_i$ actually assign a language to each state of the automaton. To find out
the behaviour of a different consequence, say $A_1\sqsubseteq B_1$, one does not need to recompute the automaton and the
functions $\wt_i$, but needs to find $\lim_{n\to\infty}\wt_n(A_1\sqsubseteq B_1)$. In other words, finding these functions provides
enough information for computing the provenance monomials of all possible consequences (in normal form) from the TBox.

In general, the construction of $\wt_i$ will not yield the fixpoint after finitely many applications. Indeed, w.r.t.\ TBox 
$\{(A\sqsubseteq B, u),(B\sqsubseteq A,v)\}$, we get that $\lim_{n\to\infty}\wt_n(A\sqsubseteq B)=(uv)^*u$, but each 
$\wt_i(A\sqsubseteq B)$ contains finitely many words. However, recall that we are not interested in the language 
$\|\Amc_{A_0\sqsubseteq B_0}\|$ \emph{per se}, but rather in the monomials that the words in this language represent.
Since the Trio semiring (which we use to characterise the provenance) uses a commutative and idempotent product operation,
we are only interested in the \emph{symbols} that appear in the words, and not in the actual words themselves. That is, we 
are only interested in the languages up to representative monomials.

\begin{definition}[\semiring-equivalence]
Two languages $\Lmc,\Lmc'$ are \emph{\semiring-equivalent} (denoted as $\Lmc\equiv_\semiring\Lmc'$) iff 
$\{[\omega]\mid \omega\in\Lmc\}=\{[\omega]\mid \omega\in\Lmc'\}$.
\end{definition}
For example, $(uv)^*u$ and $\{uv,u\}$ are \semiring-equivalent. 
While the languages $\wt_i(q)$ and the words therein may grow indefinitely, their representative monomials are limited by the
provenance variables appearing in \Tmc; which are at most $|\Tmc|$. Hence, there exists an $n\in\mathbb{N}$ such that 
$\wt_m(q)\equiv_\semiring \wt_n(q)$ holds for all $q\in Q$ and all $m\ge n$. Following Theorem \ref{thm:aut:prov}, for this $n$
$\wt_n(A_0\sqsubseteq B_0)$ contains representatives for all the provenance monomials for $A_0\sqsubseteq B_0$.

As argued before, $\wt_i(q)$ contains the weights of all runs of height at most $i$ with root $q$. It can be seen that for every run
$\rho$ of height greater than $|Q|\cdot|\Tmc|$ there is a smaller run $\rho'$ such that $\wt(\rho)=\wt(\rho')$. This means that the
least fixpoint for $\wt_i$ is found after at most $|Q|\cdot|\Tmc|$ iterations, which is polynomial in $|\Tmc|$. Specifically, the number
of iterations needed to reach a fixpoint is bounded by $\Omc(|\Tmc|^4)$.

Recall that each $\wt_i(q)$ is a language. By construction, it is a regular language; indeed, it is formed by concatenation and union
of finite languages. If we tried to represent these languages \emph{extensionally}, enumerating all the words they contain, we would
potentially need exponential space: potentially, the language may contain exponentially many words. Exploiting the
fact that these languages are regular, we can represent them through NFAs. In fact, $\wt_0$ is composed of very simple automata
with at most two states, and the construction of $\wt_{i+1}$ from $\wt_i$ requires only concatenation and union of automata,
which are basic automata operations \cite{HoUl-79automata}. However, iteratively constructing these NFA as in the 
definition of $\wt_i$ can also lead to an exponential blowup; for an example see Appendix \ref{app:computation}.
To keep the construction tractable, we exploit the succinctness power of ARAs.

Note once again that each $\wt_0(q)$ contains either a word of length 1, the empty word, or is the empty language. All these languages
are recognisable by NFA with at most two states. We call these automata $\Amc_0^q$. For each successive $\wt_{i+1}(q)$ 
we construct an automaton $\Amc_{i+1}^q$ that calls the automata $\Amc_{i}^{q'}$, which accept the languages $\wt_i(q'), q'\in Q$. 
Thus we are constructing an ARA with the ordering $\Amc_i^q\le \Amc_j^{q'}$ for all $q,q'\in Q$ and all $0\le i<j$.
Importantly, each automaton $\Amc_{i+1}^q$ requires at most five states (to concatenate the languages of the successive states)
for each transition $(q,q_1,\ldots,q_5)\in T$ (recall Table \ref{tab:transA}). Since the number of such transitions is bounded by
$|Q|^5$, it follows that the size of each ARA $\Amf_i^q:=\{\Amc_j^{q'}\mid q'\in Q, j<i\}\cup\{\Amc_i^q\}$ is in $\Omc(|Q|^5\cdot i)$. 
Let now $\Amf^q:=\Amf_n^q$, where $n$ is the number of iterations needed to reach a fixpoint w.r.t.\ \semiring-equivalence. As
seen, its size is in $\Omc(|Q|^6\cdot |\Tmc|)$; that is, it is bounded by a polynomial on $|\Tmc|$. Moreover, this ARA 
$\Amf^q$ suffices to find all the provenance monomials for the consequence $q$, as expressed next.

\begin{theorem}
\label{thm:ARAcorrect}
$\Tmc\models (A_0\sqsubseteq B_0, m)$ iff there is a word $\omega\in \Lmc(\Amf^{A_0\sqsubseteq B_0})$ such that $[m]=[\omega]$.
\end{theorem}
\begin{example}
Consider again the TBox from Example \ref{exa:treerun}.
The languages $\wt_i$ are extensionally represented in Table \ref{tab:exabehaviour}. 
\begin{table}[tb]
\caption{Extensional description of the languages $\wt_i$ for the TBox from Example \ref{exa:treerun}.}
\label{tab:exabehaviour}
\resizebox{\textwidth}{!}{
\begin{tabular}{@{}l*{10}{@{\ \ }c}@{}}
%\diagbox[width=2em]{$i$}{$q$}
& $\Box$
& $B\sqcap C\sqsubseteq D$ & $\top\sqsubseteq B$ & $A\sqsubseteq C$ & $A\sqsubseteq \exists R$ & $\exists R.B\sqsubseteq B$ 
	& $A\sqsubseteq\top$ 
	& $A\sqsubseteq B$ & $C\sqsubseteq D$ & $A\sqsubseteq D$ \\
\midrule
$\wt_0$ & $\{\varepsilon\}$ & $\{u\}$ & $\{v\}$ & $\{w\}$ & $\{x\}$ & $\{y\}$ & $\{\varepsilon\}$
	& $\emptyset$ & $\emptyset$ & $\emptyset$ \\
$\wt_1$ & \color{gray}$\{\varepsilon\}$ & \color{gray}$\{u\}$ & \color{gray}$\{v\}$ & \color{gray}$\{w\}$ & \color{gray}$\{x\}$ & \color{gray}$\{y\}$ 
	& \color{gray}$\{\varepsilon\}$
	& $\{v,xvy\}$ & $\{uv\}$ & \color{gray}$\emptyset$ \\
$\wt_2$ & \color{gray}$\{\varepsilon\}$ & \color{gray}$\{u\}$ & \color{gray}$\{v\}$ & \color{gray}$\{w\}$ & \color{gray}$\{x\}$ & \color{gray}$\{y\}$ 
	& \color{gray}$\{\varepsilon\}$
	& \color{gray}$\{v,xvy\}$ & \color{gray}$\{uv\}$ & $\{wuv,vwu,xvywu\}$ \\
$\wt_3$ & \color{gray}$\{\varepsilon\}$ & \color{gray}$\{u\}$ & \color{gray}$\{v\}$ & \color{gray}$\{w\}$ & \color{gray}$\{x\}$ & \color{gray}$\{y\}$ 
	& \color{gray}$\{\varepsilon\}$
	& \color{gray}$\{v,xvy\}$ & \color{gray}$\{uv\}$ & \color{gray}$\{wuv,vwu,xvywu\}$ \\
\end{tabular}
}
\end{table}
The construction of the automata $A_{i+1}^q$ for $i\ge 0$ and $q\in\{C\sqsubseteq D,A\sqsubseteq B,A\sqsubseteq D\}$ is
depicted in Figure \ref{fig:ARAconst}, where each transition $\m_i^q$ is a call to the automaton $\Amc_i^q$.
\begin{figure}[tb]
\centering
\includegraphics[width=\textwidth]{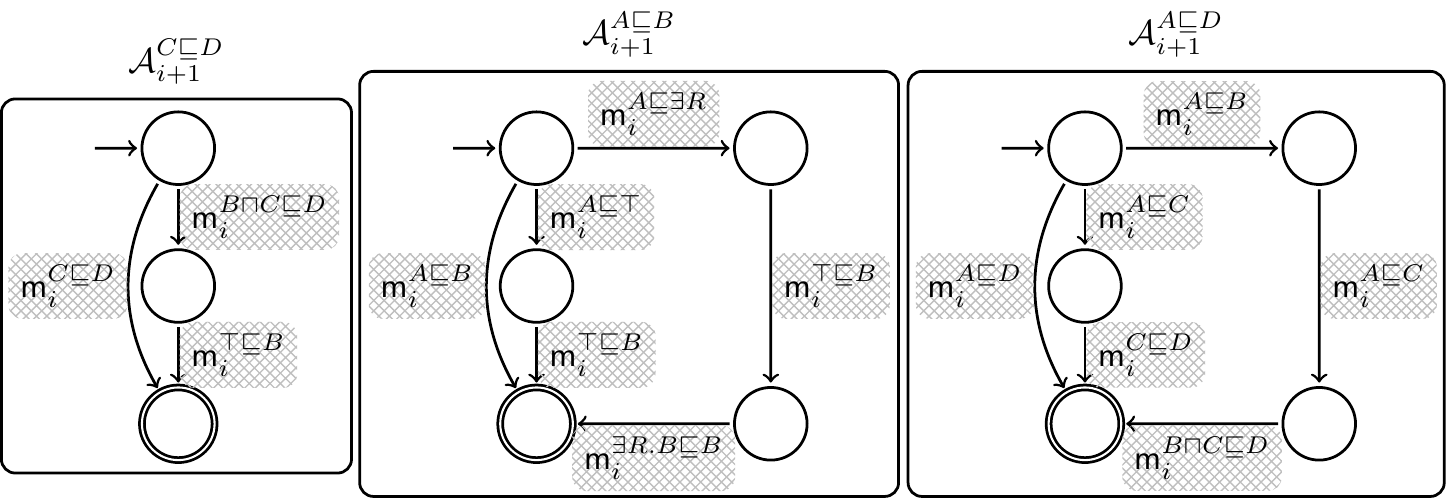}
\caption{Automata for computing $\|\Amc_{A\sqsubseteq D}\|$ w.r.t.\ the TBox in Example \ref{exa:treerun}.}
\label{fig:ARAconst}
\end{figure}
Hence, for instance $\Amc_1^{C\sqsubseteq D}$ may non-deterministically call $\Amc_0^{C\sqsubseteq D}$ (which yields the
empty language), or concatenate a word accepted by $\Amc_0^{B\sqcap C\sqsubseteq D}$ with a word from
$\Amc_0^{\top\sqsubseteq B}$. Thus, $\Lmc(\Amc_1^{C\sqsubseteq D})=\{uv\}$. Similarly, we can see that 
$\Lmc(\Amc_1^{A\sqsubseteq D})=\emptyset$.
Note that for a fixed $q\in Q$ the structure of the automata $\Amc_{i+1}^q$ is the same for all $i\ge 0$. The difference is that 
they call the automata from the previous iteration.
\end{example}
Recall that deciding whether a word $\omega$ is a accepted by an ARA \Amf is polynomial on the size of \Amf. In particular, for
$\Amf^q$ this task is polynomial on $|\Tmc|$. However, Theorem \ref{thm:ARAcorrect} requires to first find the word $\omega$
that needs to be tested. One idea is to build an automaton \Amc that accepts the language $\Lmc_m:=\{\omega\mid [\omega]=[m]\}$
and check whether $\Lmc(\Amf^q)\cap \Lmc_m\not=\emptyset$. However, it is not at all clear whether $\Lmc_m$ is even a regular
language; specifically, to the best of our knowledge it has never been verified whether the commutative closure of a regular language
it also regular. 

To solve this issue, we first \emph{guess} (in polynomial time on the size of $m$) an ordering of the symbols in $m$---say 
$\sigma_1,\ldots,\sigma_k$---and then verify
whether $\Amf^q$ accepts a word from 
\begin{align}
\sigma_1^+\sigma_2(\sigma_1\cup\sigma_2)^*\sigma_3(\sigma_1\cup\sigma_2\cup\sigma_3)^*\sigma_4\cdots
	(\bigcup_{i=1}^{k-1}\sigma_i)^*\sigma_k(\bigcup_{i=1}^{k}\sigma_i)^*;
\label{eqn:language}
\end{align}
that is, a word where the symbols first appear in the specified order. Note that the language in Equation \eqref{eqn:language}
is regular, and can be recognised by an NFA with $k+1$ states. Recall also that given an ARA \Amf and an NFA \Amc, it is possible
to construct an ARA $\Amf'$ of size bounded by $|\Amf||\Amc|$ such that $\Lmc(\Amf')=\Lmc(\Amf)\cap\Lmc(\Amc)$. Thus, verifying
whether the chosen order yields a word accepted by $\Amf^q$ is polynomial on $|\Tmc|$ and $|m|$. The non-deterministic 
ordering guess yields the following.

\begin{theorem}
Deciding $\Tmc\models (A_0\sqsubseteq B_0, m)$ is in \textsc{NP}.
\end{theorem}

\section{The Non-commutative Case}\label{sec:non-commutative}

We now consider the case where the semiring is not commutative.
The idea of non-commutativity is to preserve
the information of the order in which axioms were used to derive
a consequence. We consider here a \emph{left-absorbing product}: for multiple occurrences of
the same provenance symbol, we take into account the first (or left-most) one.
Thus, e.g., $[uvu]=[uv] \neq [vu]$.
%In this section, 
We call this case \emph{non-commutative \ELHr}.

\begin{example}
Let $\Tmc:=\{(A\sqsubseteq B,m),(B\sqsubseteq C,n)\}$.
In non-commutative \ELHr, $\Tmc\models (A\sqsubseteq C,mn)$
but  $\Tmc\not\models (A\sqsubseteq C,nm)$. 
\end{example}
Non-commutativity also means that, e.g., the concept expression 
$A\sqcap B$ is not interpreted in the same way as $B\sqcap A$,
which may seem counterintuitive since  in classical DL semantics
these concepts are equivalent.
One possible use case for this semantics is for representing definitional sentences
in natural language processing~\cite{Petrucci:2016:OLD:3092960.3092992,DBLP:conf/aime/MaD13}, %\todo{add ref}
where the order of the words usually also changes the meaning.
For example, the logic would distinguish ${\sf White}\sqcap {\sf Wine}$ from 
${\sf Wine}\sqcap {\sf White}$.
  
Interestingly, we know from Equation \eqref{eqn:language} that we can verify whether $\Amf^q$ accepts a representative
(under left-absorption) of a monomial $m$. The benefit in this case is that it is not necessary to first guess the right ordering,
as it is required by the ordering given in $m$. This yields the following result.

\begin{theorem}
$\Tmc\models (A_0\sqsubseteq B_0, m)$ w.r.t.\ a left-absorbing, non-commutative semiring can be decided in polynomial time.
\end{theorem}

\section{Conclusions}

In this paper we have studied the complexity of deciding whether the provenance of a subsumption relation contains a given
monomial $m$. In previous work \cite{DBLP:conf/ijcai/BourgauxOPP20}, it was shown through a completion algorithm, that this problem 
is in \textsc{PSpace} when the semiring product is idempotent and commutative, but only a polynomial lower bound (derived from 
reasoning in \ELHr) was given. By viewing the completion
algorithm backwards, as a decomposition approach based on tree automata, and exploiting a less known class of automata (ARAs)
to simulate structure sharing, we were able to lower this upper bound to \textsc{NP}. Unfortunately, the polynomial lower bound
remains the best available at the moment. If we substitute commutativity by a notion of left-absorption, we obtain a tight polynomial-time
complexity for this problem. Interestingly, the technique developed can be applied to instance queries and assertion entailments,
simply by extending the automaton construction to the ABox-handling rules from \cite{DBLP:conf/ijcai/BourgauxOPP20}. Hence,
the same complexity bounds hold in both cases. One avenue for future work is to close the remaining complexity gaps in these
problems.

Note that the complexity results depend strongly on the properties of the provenance semiring. Indeed, the fixpoint computation
of the functions $\wt_i$ only terminates due to the idempotence and commutativity (or left-absorption) of the product. However,
the construction of the automaton $\Amc_{A_0\sqsubseteq B_0}$ remains correct for a larger class of semirings (Theorem
\ref{thm:aut:prov}). We will study whether the technique can be applied in practice for these other semirings. One particular point
of interest is to consider \emph{closure semirings} \cite{Lehm-TCS77} or other approaches for handling the repeating structure
of the ARAs.

\bibliographystyle{splncs04}
\bibliography{references}

%\backmatter
\newpage

\appendix

\section{Examples}

In this appendix we provide some examples which should help further understand the notions and constructions introduced in the
paper, along with relevant properties of ARAs.

\subsection{ARA Succinctness}
\label{app:succ}

We start by considering the question of succinctness of ARAs. For this, consider the language $\Lmc_n$ containing only the word 
$a^{2^n}$; that is, the symbol $a$ repeated $2^n$ times. Since this language is finite (it contains only one word) it is also regular. 
Moreover, any NFA that accepts $\Lmc_n$ must have at least $2^n$ states. We show that this language can be accepted by an 
ARA of size polynomial on $n$.

\begin{figure}[b]
\centering
\includegraphics[width=0.48\textwidth]{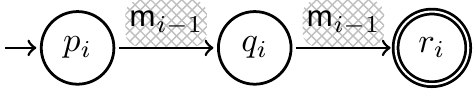}
\caption{The automata $\Amc_i$, $i>1$. $\Amc_1$ is similar, but transitions read the symbol $a$.}
\label{fig:expARA}
\end{figure}

Given a fixed $n\in\mathbb{N}$ we define the automata $\Amc_i=(Q_i,\Sigma_i,\Delta_i,I_i,F_i)$ for $1\le i\le n$ with the usual ordering
over natural numbers, where:
\begin{itemize}
\item $Q_i=\{p_i,q_i,r_i\}$;
\item $\Sigma_i=\{a\}\cup \{\m_j\mid 1\le j<i\}$;
\item $\Delta_1=\{(p_1,a,q_1),(q_1,a,r_1)\}$, and $\Delta_i=\{(p_i,\m_{i-1},q_i),(q_i,\m_{i-1},r_i)\}$ for $1<i\le n$;
\item $I_i=\{p_i\}$; and
\item $F_i=\{r_i\}$.
\end{itemize}
These automata are depicted in Figure \ref{fig:expARA} for $i>1$. 
As it can be seen, $\Amc_i$ first calls the automaton $\Amc_{i-1}$
and then calls that same automaton again. Thus $\Amc_i$ accepts the concatenation of $\Amc_{i-1}$ with itself. The graphical 
representation of $\Amc_i$ is identical, but each transition reads the symbol $a$ instead. Thus, $\Lmc(\Amc_1)=\{aa\}$. It can be
seen, through a simple inductive argument, that for each $i$, $\Lmc(\Amc_i)=\{a^{2^i}\}$. Hence, the ARA 
$\Amf_n=\{\Amc_i\mid 1\le i\le n\}$ accepts $\Lmc_n$. Moreover, $\Amf_n$ has in total $3n$ states (three for each automaton
$\Amc_i$).

\begin{figure}[tb]
\includegraphics[width=\textwidth]{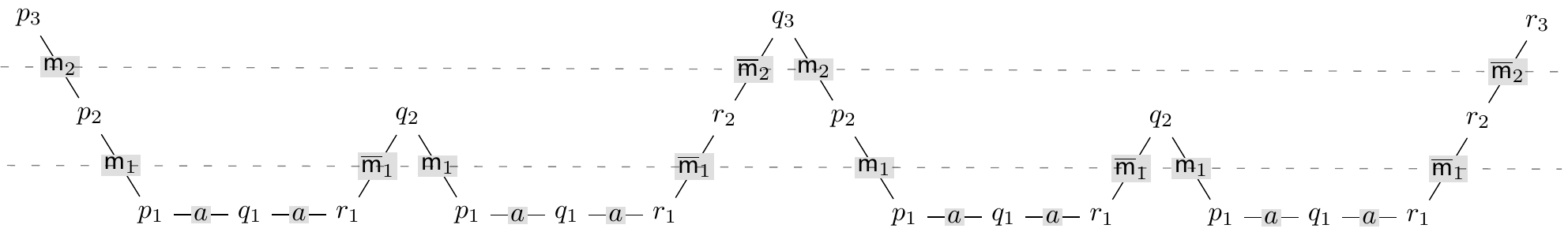}
\caption{The only successful run for the automaton $\Amf_3$.}
\label{fig:run}
\end{figure}

Note that these automata are all deterministic, and they all accept exactly one word. Hence, there is exactly one successful 
run for each of them. For the sake of the example, let us consider $\Amf_3$, whose language is the singleton $\{a^8\}$. The
successful run for this automaton is depicted in Figure \ref{fig:run}.
We visually separate the execution of each automaton $\Amc_i$ by a dashed line.  Note that $\Amc_3$ is only responsible 
for calling $\Amc_2$ twice (through the $\m_2$ transition) and receive the signal that the execution of $\Amc_2$ is finished
through the $\overline\m_2$ transition). $\Amc_2$ behaves similarly, just calling $\Amc_1$ twice; hence, $\Amc_1$ is called a 
total of four times---twice for each call to $\Amc_2$. In the end, $\Amc_1$ is a simple deterministic automaton, which reads the
symbol $a$ twice. By removing the call transitions $\m_i, \overline\m_i$ from the run, we obtain that the word accepted by this
run is exactly $a^8$, as expected.

\subsection{Behaviour Computation}
\label{app:computation}

We now exemplify the cases where the behaviour computation may suffer from an exponential blowup, along with the benefits of
the structure sharing-like strategy introduced by ARAs.

\begin{figure}[b]
\includegraphics[width=\textwidth]{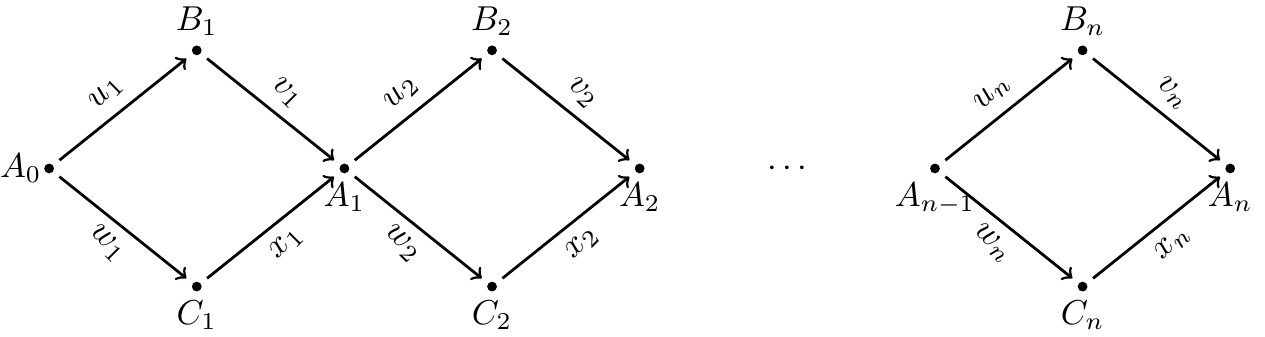}
\caption{The TBox $\Tmc_n$, or the minecraft sword.}
\label{fig:Tn}
\end{figure}

Given $n\in\mathbb{N}$, let $\Tmc_n$ be the annotated TBox defined by
\begin{align*}
\Tmc_n := \{ & (A_{i-1} \sqsubseteq B_i, u_i), (A_{i-1} \sqsubseteq C_i, w_i), (B_i \sqsubseteq A_i, v_i), (C_i \sqsubseteq A_i, x_i)
	\mid 1\le i\le n\}.
\end{align*}
This very simple TBox is depicted in Figure~\ref{fig:Tn}.
It is a simple exercise to verify that $\Tmc_n\models (A_0\sqsubseteq A_n, m)$ iff $[m]=[m_1m_2\cdots m_n]$ for a selection
of $m_i\in\{u_iv_i,w_ix_i\}$. This means that the provenance polynomial for this consequence has $2^n$ monomials; one for each
choice of the elements $m_i$. Thus, if $\Amc_i^{q}$ are the ones constructed in Section \ref{sec:behaviour}, we get that 
$\Lmc(\Amc_n^{A_0\sqsubseteq A_n})$ contains $2^n$ words, and hence cannot be described extensionally in only polynomial
space (or time). 

The attentive reader will notice that $\Lmc(\Amc_n^{A_0\sqsubseteq A_n})$ is recognised by an NFA with $3n+1$ states; in fact,
this automaton would have an uncanny resemblance to Figure \ref{fig:Tn}, reading the elements as states. 
Unfortunately, reaching this small NFA automatically
requires (at least) a minimisation step, which is hard in general \cite{JiRa-SIAM93}. Indeed, a direct construction of the automata
$\Amc_i^q$ using the definition from Section \ref{sec:behaviour} would yield automata of exponential size, as described next.

Recall that by construction, $\wt_{i+1}(q)=\wt_i(q)\cup \bigcup_{(q,q_1,q_2)\in T}\wt_i(q_1)\cdot \wt_i(q_2)$.%
\footnote{In reality, we would look at the trees of arity 5, but since this TBox uses only atomic concept inclusions, we focus on the
binary tree constructed by this case only.}
In particular, the unions in $\wt_{i+1}(q)$ for $q\in{A_0\sqsubseteq A_k, A_0\sqsubseteq B_k, A_0\sqsubseteq C_k}$, $k\ge 1$ 
contain (among others) the following concatenations respectively:
\begin{description}
\item[[$A_0\sqsubseteq A_k$\negmedspace]] $\wt_i(A_0\sqsubseteq B_k)\cdot \wt_i(B_k\sqsubseteq A_k)$; 
		$\wt_i(A_0\sqsubseteq C_k)\cdot \wt_i(C_k\sqsubseteq A_k)$;
\item[[$A_0\sqsubseteq B_k$\negmedspace]] $\wt_i(A_0\sqsubseteq A_{k-1})\cdot \wt_i(A_{k-1}\sqsubseteq B_k)$;
\item[[$A_0\sqsubseteq C_k$\negmedspace]] $\wt_i(A_0\sqsubseteq A_{k-1})\cdot \wt_i(A_{k-1}\sqsubseteq C_k)$.
\end{description}
Following the standard construction for the concatenation and union of automata, the automaton for 
$\wt_{i+2}(A_0\sqsubseteq A_k)$ would thus have copies of each of the automata for $\wt_{i+1}(q)$ with 
$q\in\{A_0\sqsubseteq B_k, B_k\sqsubseteq A_k,A_0\sqsubseteq C_k, C_k\sqsubseteq A_k\}$. But note that
$\wt_{i+1}(A_0\sqsubseteq B_k)$ and $\wt_{i+1}(A_0\sqsubseteq C_k)$ both contain a copy of
$\wt_{i}(A_0\sqsubseteq A_{k-1})$. That is, $\wt_{i+2}(A_0\sqsubseteq A_k)$ contains (at least) two copies of 
$\wt_{i}(A_0\sqsubseteq A_{k-1})$. Since this holds true for all $k\ge 1$, it immediately follows that $\wt_{2n+2}(A_0\sqsubseteq A_k)$
contains $2^n$ copies of $\wt_2(A_0\sqsubseteq A_1)$, and hence at least $2^n$ states.

Note that construction using ARAs avoids this exponential explosion by abstaining from making copies of the automata at previous
levels, but calling them several times instead.

\end{document}